\documentclass[10pt,twocolumn,twoside]{IEEEtran}

\usepackage{psfrag,epsfig,graphics}
\usepackage{amsmath,amssymb,amsthm,multirow}
\usepackage{cite,subfigure,balance}
\usepackage[labelsep=period]{caption}
\usepackage{algorithm}
\usepackage{algorithmic}
\usepackage{setspace}
\usepackage{bm}
\usepackage{bbm}
\usepackage{cases}

\DeclareMathOperator*{\argmin}{argmin}

\usepackage{color}

\definecolor{violet}{rgb}{0.5,0,0.5}

\def\R{\ensuremath{{\mathrm{I\!R}}}}

\newcommand{\cb}[1]{{\boldsymbol{#1}}}
\newcommand{\cp}[1]{\ifmmode {\mathcal{#1}}\else ${\mathcal{#1}}$\fi}
\newcommand{\balpha}{\boldsymbol{\alpha}}

\newcommand{\bC}{\boldsymbol{C}}

\newcommand{\bI}{\boldsymbol{I}}

\newcommand{\bW}{\boldsymbol{W}}

\newcommand{\bc}{\boldsymbol{c}}

\newcommand{\bs}{\boldsymbol{s}}

\newcommand{\bw}{\boldsymbol{w}}

\newcommand{\bx}{\boldsymbol{x}}

\newcommand{\bTheta}{\boldsymbol{\Theta}}

\newcommand{\bPhi}{\boldsymbol{\Phi}}

\newcommand{\bpsi}{\boldsymbol{\psi}}

\newcommand{\bphi}{\boldsymbol{\phi}}
\newcommand{\bPsi}{\boldsymbol{\Psi}}

\newcommand{\E}{{\mathbb{E}}}

\usepackage{tikz}

\begin{document}

\title{The `D-Subspace' Algorithm for \\Online Learning over Distributed Networks}
%
\author{Yitong Chen$^\star$, \IEEEmembership{Student Member, IEEE}, \ Danqi Jin$^\star$, \ Jie Chen, \IEEEmembership{Senior Member, IEEE},\\ \ C{\'e}dric Richard, \IEEEmembership{Senior Member, IEEE}

\thanks{Y. Chen and D. Jin are co-first authors with equal contributions to this material.
Y. Chen and J. Chen are with Centre of Intelligent Acoustics and Immersive Communications at School of Marine Science and Technology, Northwestern Polytechnical University, Xi'an, China (chenyitong@mail.nwpu.edu.cn).
D. Jin is with School of Electronic Information, Wuhan University, Wuhan, China (danqijin@whu.edu.cn).
C.~Richard is with Universit\'{e} C\^{o}te d'Azur, OCA, CNRS, France.}
}

\maketitle

\textbf{Notation.} Normal font letters $x$, bold lowercase letters $\bx$ and bold uppercase letters $\cb{X}$ denote scalars, column vectors and matrices, respectively. The notation $[\,\cdot\,]_{(:,j)}$ refers to the $j$-th column. The mathematical expectation is denoted by $\E\{\cdot\}$. The set ${\cal{N}}_k$ denotes the neighbors of node $k$ (including $k$ itself), and $|{\cal N}_k|$ denotes its cardinality. Notation $\bigl[\bw_\ell\bigr]_{\ell\in{\cal N}_k}$ represents a matrix consisting of all $\bw_\ell$ with $\ell\in{\cal N}_k$.

This material introduces the D-Subspace algorithm derived on the basis of the centralized algorithm~\cite{Nassif2020sub}, which originally addresses parameter estimation problems under subspace constraint.

Consider a connected network with $N$ agents. The set of all agents is denoted as ${\cal N}\triangleq \{1,2,\cdots,N\}$. Each agent $k \in{\cal N}$ is endowed with a strongly convex, real-valued and differentiable cost function $J_k(\bw_k)$, which corresponds to the expectation of a loss function $G_k(\bw_k;\bs_{k,n})$:
\begin{equation}
\label{eq:loss-function}
J_k(\bw_k) \triangleq \E\{G_k(\bw_k;\bs_{k,n})\},
\end{equation}
where the expectation operator $\E\{\cdot\}$ is evaluated over the distribution of random data $\bs_{k,n}$, with subscripts $k$ and $n$ representing node index and time instant, respectively. We denote the real-valued parameter vector $\bw_k^\star \in{\mathbb R}^L $ as the unique minimizer of $J_k(\bw_k)$. Define a matrix $\bW^\star$ as:
\begin{equation}
\label{eq:def-W*}
\bW^\star\triangleq [\bw_1^\star, \ \ \bw_2^\star,\ \ \cdots, \ \ \bw_N^\star]\in{\mathbb R}^{L\times N}
\end{equation}
The aim of this material is to explore a situation where $\bW^\star$ is a low-rank matrix with rank $r^{\star}$. In this case, we have:
\begin{align}
\label{eq:def-w*}
\bw^\star_k&=\sum_{i=1}^{r^\star}\alpha_{k,i}^{o}\bc_{i}= \bC\cdot \balpha_k^{o}
\end{align}
where $\{\bc_{i}\}_{i=1}^{r^\star}$ is a basis, $\{\alpha_{k,i}^{o}\}_{i=1}^{r^\star}$ are corresponding weights, matrix $\bC\triangleq [\bc_{1}\, \bc_{2} \cdots \bc_{r^\star}]\in\R^{L\times r^{\star}}$, and vector $\balpha_k^{o}\triangleq [\alpha_{k,1}^{o}\, \alpha_{k,2}^{o} \cdots \alpha_{k,r^\star}^{o}]^{\top}$. In this material, we assume that $\balpha_k^{o}$ is known before. Substituting \eqref{eq:def-w*} into \eqref{eq:def-W*}, we have:
\begin{equation}
\label{eq:def-W*-new}
\bW^\star= \bC\cdot \bTheta^{o}
\end{equation}
where matrix $\bTheta^{o}\triangleq[\balpha_1^{o}\, \balpha_2^{o}\,\cdots \balpha_{N}^{o}]\in\R^{r^{\star}\times N}$ is assumed to be known. Consequently, a centralized optimization problem emerges:
\begin{align}
&\argmin_{\bw_{\ell:{\ell\in{\cal N}}}}{\sum}_{\ell=1}^NJ_\ell(\bw_\ell)\notag\\
&{\rm s.t.}\quad \bigl[\bW^\top\bigr]_{(:,j)} \in {\cal R}\Bigl([\bTheta^{o}]^{\top}\Bigr), \,\,\forall\,\,j\label{pro:global}
\end{align}
where $\bW\triangleq\bigl[\bw_\ell\bigr]_{\ell\in{\cal N}}$ is an estimate of $\bW^\star$, and ${\cal R}(\cdot)$ denotes the range space operator. In order to solve \eqref{pro:global} iteratively, the gradient projection method can be applied, resulting in:
\begin{numcases}{}
{\bpsi}_{k,n+1} = \bw_{k,n} -\mu_k\nabla_{\bw_k} G_k(\bw_{k, n};\bs_{k,n})&\label{eq:DistributedProjectedGD-Centralized}\\
{\bPsi}_{n+1} \triangleq [{\bpsi}_{1,n+1}, \,{\bpsi}_{2,n+1},\, \cdots, {\bpsi}_{N,n+1}]&\label{eq:Psi-estimate}\\
\bPhi_{n+1} =\bigl[{\cal P}_{[\bTheta^{o}]^\top}\cdot({\bPsi}_{n+1}^{\top})\bigr]^{\top}={\bPsi}_{n+1}\cdot{\cal P}_{[\bTheta^{o}]^\top}&\label{eq:DProjectedSGD-Centralized}
\end{numcases}
where the projection matrix ${\cal P}_{[\bTheta^{o}]^\top}$ is defined as:
\begin{align}
\label{eq:proj-1}
{\cal P}_{[\bTheta^{o}]^\top} &\triangleq [\bTheta^{o}]^\top(\bTheta^{o}[\bTheta^{o}]^\top)^{-1}\bTheta^{o}.
\end{align}
Equations \eqref{eq:DistributedProjectedGD-Centralized}-\eqref{eq:DProjectedSGD-Centralized} are centralized solution, abbreviated as `C-Subspace' in this material.

We also aim to introduce a distributed solution. Given that the network is connected and only local data exchanges are allowed, we define a local optimal matrix $\bW^\star_k$ for each node $k$ as:
\begin{equation}
\label{eq:def-W*-k}
\bW_k^\star\triangleq\bigl[\bw_\ell^\star\bigr]_{\ell\in{\cal N}_k}\in\R^{L\times|{\cal N}_k|}
\end{equation}
To ensure the uniqueness of $\bW_k^\star$, we arrange its columns~$\bw_\ell^\star$ with $\ell \in \mathcal{N}_k$ in ascending order with respect to $\ell$, such that~$\bw_\ell^\star$ is its $i_{\ell}^{(k)}$-th column.
Within the neighborhood ${\cal N}_k$ of each node $k$, we have~\footnote{Note that notation $(\cdot)^{(k)}_{\ell,\cdot}$ denotes a quantity related to node $\ell$, which is evaluated at node $k$ and provided by node $k$.}:
\begin{align}
\label{wk_basis}
\bw^\star_{\ell}=\sum_{i=1}^{r_k^\star}\alpha_{\ell,i}^{(k)}\bc_{k,i}= \bC_k\cdot \balpha_{\ell}^{(k)},\,\,\forall \ell\in{\cal N}_k
\end{align}
where $r_k^\star$ is the rank of matrix $\bW_k^\star$, $\{\bc_{k,i}\}_{i=1}^{r_k^\star}$ is a basis, with $\bigl\{\alpha_{\ell,i}^{(k)}\bigr\}_{i=1}^{r_k^\star}$ the corresponding weights with respect to node $k$, matrix $\bC_k\triangleq [\bc_{k,1}\, \bc_{k,2} \cdots \bc_{k,r_k^\star}]\in\R^{L\times r_k^{\star}}$, and vector $\balpha_{\ell}^{(k)}\triangleq \bigl[\alpha_{\ell,1}^{(k)}\, \alpha_{\ell,2}^{(k)} \cdots \alpha_{\ell,r_k^\star}^{(k)}\bigr]^{\top}$. To ensure the uniqueness of \eqref{wk_basis}, we require that for all $k$, all $\ell\in{\cal N}_k$ and all $i\in\{1,2,\cdots,r_k^\star\}$, there exists a $j\in\{1,2,\cdots,r^\star\}$, such that:
\begin{align}
\alpha_{\ell,i}^{(k)} = \alpha_{\ell,j}^{o} \,\,\, {\rm and} \,\,\,\bc_{k,i} = \bc_{j}.
\end{align}
Similarly, in this material, we assume that $\balpha_{\ell}^{(k)}$ is known before. Substituting \eqref{wk_basis} into \eqref{eq:def-W*-k}, we have:
\begin{equation}
\label{eq:def-W*-new}
\bW_k^\star= \bC_k\cdot \bTheta_k
\end{equation}
where matrix $\bTheta_k\triangleq\bigl[\balpha_{\ell}^{(k)}\bigr]_{\ell\in{\cal N}_k}\in\R^{r_k^{\star}\times|{\cal N}_k|}$ is known, with $\balpha_{\ell}^{(k)}$ being its $i_{\ell}^{(k)}$-th column. Consequently, a distributed optimization problem emerges at each node $k$ as:
\begin{align}
&\argmin_{\bw_{\ell:{\ell\in{\cal N}_k}}}{\sum}_{\ell\in{\cal N}_k}J_\ell(\bw_\ell)\notag\\
&{\rm s.t.}\quad \bigl[\bW_k^\top\bigr]_{(:,j)} \in {\cal R}\Bigl([\bTheta_{k}]^{\top}\Bigr), \,\,\forall\,\,j\label{pro:local}
\end{align}
where $\bW_k$ is an estimate of $\bW_k^\star$, with its $i_{\ell}^{(k)}$-th column serving as an estimate of $\bw_\ell^{\star}$. To solve \eqref{pro:local} iteratively, we apply the gradient projection method. Local counterparts corresponding to the same estimate are further combined~\cite{ref2}, resulting in:
\begin{numcases}{}
{\bpsi}_{k,n+1} = \bw_{k,n} -\mu_k\nabla_{\bw_k} G_k(\bw_{k, n};\bs_{k,n})&\label{eq:DistributedProjectedGD-D}\\
{\bPsi}_{k,n+1} \triangleq \bigl[{\bpsi}_{\ell,n+1}\bigr]_{\ell\in{\cal N}_k}&\label{eq:Psi-estimate-D}\\
\bPhi_{k,n+1} =\bigl[{\cal P}_{[\bTheta_{k}]^\top}\cdot({\bPsi}_{k,n+1}^{\top})\bigr]^{\top}={\bPsi}_{k,n+1}\cdot{\cal P}_{[\bTheta_{k}]^\top}&\label{eq:DProjectedSGD-D}\\
{\bphi}_{k,n+1}^{(\ell)} \triangleq \bigl[\bPhi_{\ell,n+1}\bigr]_{(:,{i}_{k}^{(\ell)})}&\\
\bw_{k,n+1} = \sum_{\ell\in{\cal N}_k}a_{\ell k}\bphi_{k,n+1}^{(\ell)}&\label{eq:combination}
\end{numcases}
where the projection matrix ${\cal P}_{[\bTheta_{k}]^\top}$ is defined as:
\begin{align}
\label{eq:proj-2}
{\cal P}_{[\bTheta_k]^\top} &\triangleq [\bTheta_k]^\top(\bTheta_k[\bTheta_k]^\top)^{-1}\bTheta_k,
\end{align}
and $a_{\ell k}$ are single-task combination coefficients satisfying:
\begin{align}
    \sum_{\ell\in{\cal N}_k}a_{\ell k}=1,\;a_{\ell k}\geq 0,\; {\rm and}\; a_{\ell k}=0\; {\rm{if}}\;\ell\notin {\cal{N}}_k. \label{eq:A-Constraints}
\end{align}
Moreover, in several cases, the diagonal loading technique can be incorporated into \eqref{eq:proj-2}, resulting in:
\begin{align}
\label{eq:proj-3}
{\cal P}_{[\bTheta_k]^\top} &\triangleq [\bTheta_k]^\top(\bTheta_k[\bTheta_k]^\top + \eta\bI)^{-1}\bTheta_k,
\end{align}
where $\eta>0$ is diagonal loading factor with a small value. Equations \eqref{eq:DistributedProjectedGD-D}-\eqref{eq:combination} are distributed solution, abbreviated as `D-Subspace' in this material.


\end{document}